# The Making of 5G – Building An End-to-End 5G-Enabled System

Idelkys Quintana-Ramirez, Anthony Tsiopoulos, Maria A. Lema, Fragkiskos Sardis, Luis Sequeira, James Arias, Aravindh Raman, Ali Azam, Mischa Dohler *FIEEE FREng FRSA*

Centre for Telecommunications Research
Department of Informatics, King's College London (KCL), UK

**Abstract:** This article documents one of the world's first standards-compliant pre-commercial end-to-end 5[th] generation (5G) systems. Focus is on a standardized 5G architecture which includes the underlying 3GPP components but also the ETSI Network Function Virtualization (NFV) management and orchestration capabilities. The truly innovative character of 5G enabling fundamental changes to architecture and implementation is discussed, and details of monitoring and orchestration approaches that are deemed instrumental in unlocking the full potential of 5G. Finally, it is important to us to share the lessons learned which we hope are of use to industry and academia alike when building, deploying and testing emerging 5G systems.

## 1. Introduction

When plotting cellular Key Performance Indicators (KPIs) in semi-logarithmic scale over time, a surprising trend can be observed: the KPIs improve by one or several orders of magnitude as new generations emerge. Specifically, in the context of the transition from 4G to 5G [1], that means that an air interface peak data rate of 1Gbps in 4G will need to yield 10Gbps or more in 5G. An air interface latency of about ten milliseconds in 4G needs to come down to one millisecond or less in 5G. The number of devices able to attach to a single base station will increase from tens of thousands of devices to several hundred thousand devices.

Interestingly, the KPIs are so performant in 5G that the system is not only of great utility to consumers but – for the first time – also to industries. It is exciting to see that 5G, a standardized general-purpose platform technology, will be able to replace expensive, often proprietary, industry systems in emerging Industry 4.0 applications. 5G will be able to support ultra-low latency and ultra-reliable control applications and industrial Internet of Things (IoT) applications, and thereby aid efficiency and effectiveness of various industrial verticals.

The technical realization of such a versatile communications system, however, requires some very transformative changes to ensure that it remains competitive from a cost and performance point of view. Flexibility and openness have traditionally proven to be the most important design parameters to ensure scale, competitiveness and adoption. A development within 5G is hence the ability to be much more flexible, which is ensured through the following three important design changes.

First, in the 5G radio access and core networks, hardware and software are becoming clearly separated from each other [2]. It means that 5G features are virtualized in software and delivered over commodity hardware where it runs as virtual machines or containers. Such a decoupling is important as it enables each ecosystem to innovate independently and at their respective pace. It has proven very successful in the computing industry, where hardware (computer), middleware (operating system) and software (applications) are developed independently. Underpinned by our own design and deployment experiences, we are confident that it will also prove highly successful in the telecoms ecosystem.

Second, within the software, we now observe a much stronger atomization of functionalities, not least underpinned by a clear separation between data and control plane [3]. The clear separation of software functionalities allows us to potentially replace certain functionalities much quicker with more advanced embodiments. Therefore, incremental improvements of the technology can now occur more easily and without having to change physical devices or firmware.

Third, the atomized software components can be much easier virtualized, then arranged and placed as per need. Advanced functionalities can thus be moved flexibly, resources instantiated in a moment and IoT services delivered at scale [4].

The above design changes will trigger new waves of innovation within the telco architecture as well as for over-the-top services. This however bears new challenges, namely in terms of how to manage such a flexible infrastructure with such a high design degree of freedom and many resulting conflicting situations. In addition, how to ensure that the flexibility is standardized. And, most importantly, how to demonstrate its viability in terms of practical deployments and trials.

The focus of this paper is thus to detail a large 5G deployment comprised of different scenarios for testing which relies on above design principles. The paper is structured as follows: we first describe the standardized 5G end-to-end architecture which includes the underlying 3GPP components but also the ETSI NFV management and orchestration capabilities. In Section 3, we describe in detail the implementation and functioning of ETSI's orchestration framework and how it can manage the complex ecosystem. We then discuss specific topology, integration and deployment challenges which we faced and which we deem important to share so as to facilitate smooth 5G IoT and broadband delivery in the future. Finally, we conclude with a summary of our outlook.

## 2. End-to-End 5G-Enabled Elements and Architecture

The deployed end-to-end system consists of multiple 5G-enabled components as well as the integration of legacy system architectures. All network functions are fully virtualized and run in virtual machines in different cloud infrastructures, which are all managed and orchestrated using a common orchestration framework. The systems integrated are a combination of pre-commercial prototype 5G equipment, commercial 4G equipment and experimental implementations, integrating the latest contributions in network convergence as well as cloud based solutions for mobile architectures, including Cloud Radio Access Network (RAN) and core network functional splits. The focus of this work is to allow the seamless integration of such a heterogeneous system, as shown in Fig. 1. All this, using open source cloud technologies, and contributing to the state-of-the-art of network orchestration and management.

The 5G system consists of a set of outdoor sites, transmitting at 5G pioneering bands, 3.5GHz and 28GHz, allowing for increased throughputs and lower latencies in the access network. The virtualized RAN and core network functions are running in separate cloud environments, which are interfaced with the management and orchestration framework; the reader is referred to the subsequent section for more in-depth explanation of the control, management and orchestration framework.

The 4G commercial system considers a set of indoor small cells, transmitting at 700MHz, with Narrow-Band-IoT (NB-IoT) enabled capabilities, as defined by 3GPP in [5]. The 4G core network is virtualized and supports end-to-end network slicing, allowing multiple instantiations of Virtualized Network Functions (VNF) to satisfy different levels of QoS and service isolation. Similarly, the 4G virtualized core is interfaced with the orchestrator.

On the experimental side, we have integrated specific network services implementation elements resulting from multiple research works in the context of IoT, access management and core network function placement. In the following, we describe these elements in more detail.

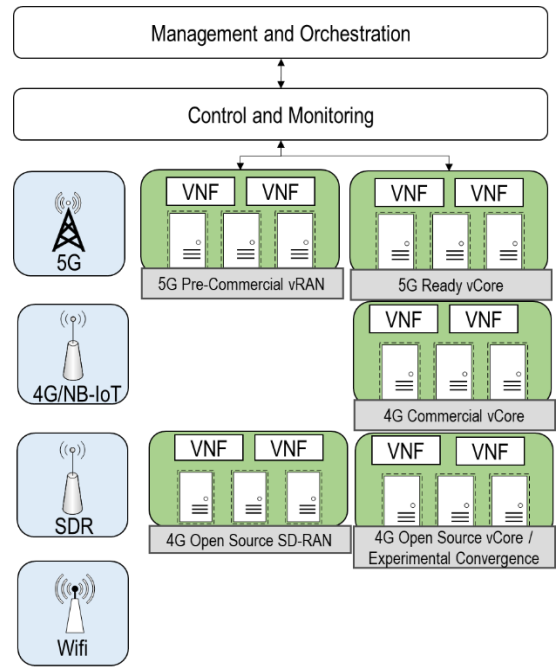

*Figure 1:* High Level Testbed Summary of included Access Technologies, Access Networks, Core Network and Orchestration.

## 2.1 Cloud RAN Functional Split Integration

Virtualization in the context of the access network has gained a lot of interest in both research community and standardization bodies [6,7]. In such context, there is no need to have the entire access protocol stack co-located with the transmission part, and more flexible deployments of the access network can be considered. Particularly interesting is the separation of the medium access control layer functions from the physical layer ones, as it allows for more cooperative techniques in the radio resource management. Multiple research studies have attempted to find the optimal functional split in the RAN, which maximizes capacity by enabling cooperation between different transmitters as well as reducing the stress in the transmission fronthaul network. However, it has been concluded that there is no rule of thumb, and several variables should be considered when assessing the optimal deployment [6]. In the context of virtualization and orchestrated networks, the selection of optimal deployment based on a case-to-case basis is possible through the instantiation of multiple network services.

The integrated cloud RAN implementation allows for different functional splits across the RAN protocol stack, which constitute different network services and which can be deployed and instantiated based on the use case requirements. For instance, in high-bandwidth IoT applications, cloud RAN capabilities are instrumental in ensuring a viable trade-off between performance and delay.

## 2.2 Virtualized Core Network Split

Like the RAN, the virtualization of core network functions allows more flexible deployments which can address some of the KPIs required for 5G paving way towards a service-oriented core. Specifically, 3GPP considers a more modular core network architecture for 5G [8], where the control and user plane functions are completely decoupled and communicate with each other through new interfaces.

Such modularity and clear separation of network functions enables new and innovative deployment options, which can facilitate edge computing capabilities [9], as shown in Fig. 2.

When control and user plane functions are separated, the user plane, which operates at a more stringent time scale than the control plane, can reside closer to the edge as a local breakout for content and service provisioning. Such a deployment allows the decentralization of services and distribution of content caching across the network which addresses both latency and congestion in the transport network.

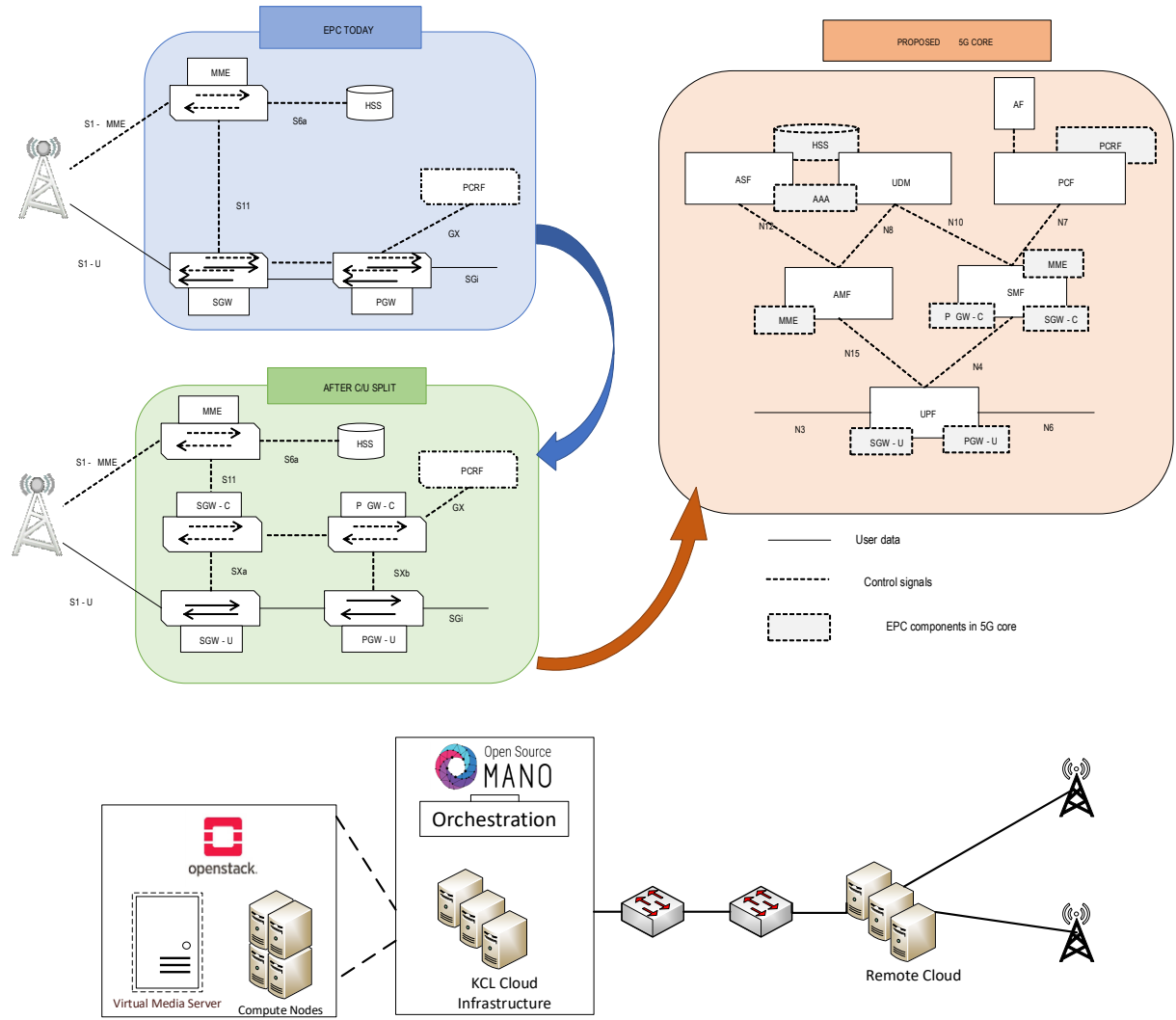

*Figure 2: Remote Cloud Implementation to support flexible network deployments and edge computing.*

**2.3 Capitalization of Heterogeneity through Multiple Access**

The 5G network architecture is expected to act as an integrator of multiple access networks: from cellular access (i.e., 3GPP networks) to multiple radio access (i.e., Wi-Fi or IoT-specific) and fixed access. To capitalize the number of resources available and to maximize the use of all available networks, the 5G architecture is intended to support the simultaneous access across different technologies. This is known as network convergence, and both 3GPP and the Broadband Forum have worked on enabling architectures to support it [10,11].

In the presented deployment, we aimed to replicate that architecture by including an Hybrid Access Gateway (HAG) functionality which acts as a proxy and enables traffic to flow to the user via multiple heterogeneous networks. We implemented a Multi-Path TCP (MPTCP) VNF which capitalizes on the 5G network deployment. This specific MPTCP implementation aggregates traffic at the HAG and allows for path selection based on a set number of parameters that can be used to optimize resource allocation. In this embodiment, the UE can switch between network interfaces that send and receive data from the HAG. The HAG includes the traffic scheduling and appropriate polices to be enforced, which are complemented with the monitoring and control parts of the orchestration framework to ensure a satisfactory end-to-end delivery.

## 3. Cloud Orchestration

Since the inclusion of the NFV framework, all network functions included in the communications systems are a combination of physical elements and software that runs in cloud infrastructures. Cloud technologies are therefore a critical tool to allow a dynamic deployment and management of these VNFs. One of the requirements for the cloud infrastructure is that it must accommodate different technologies, often part of legacy systems, such as 4G, and the integration of heterogeneous systems, such as the diverse IoT ecosystem. A key part of this platform is to provide mechanisms for management and orchestration of multiple systems, and ensure interoperability across all. To this end, we explore both cloud and orchestration solutions that allow to expand its capabilities to control multiple network services. For the cloud architecture considered in this paper, the following design premises should be included:

- A cloud operating system controls compute, storage and networking resources through a datacenter.
- An orchestrator provides the necessary framework to achieve seamless integration of heterogeneous systems.
- The Virtual Infrastructure Management (VIM) manages the VNFs based on VMs and containers.
- A SDN solution provides the capability to dynamically program traffic flows to modify virtual local area network based traffic matching rules.

Usually, each VIM has its own set of management tools which provides a web-based user interface to control VIM's services (for instance, Horizon is the module in charge of such as task in OpenStack). However, the management is within the corresponding VIM and could potentially become meaningless in multi-VIM environments mainly when different vendors are involved. Multiple orchestration solutions have emerged in the last few years: OpenBaton, Cloudify, Kubernetes (which has been discarded because it only orchestrates containers) and Open Source MANO (OSM) that is provided by ETSI. The Management and Orchestration (MANO) component is addressed via OSM, a software stack that enables the orchestration, synchronization and lifetime management of VNFs or network services. OSM have been selected as our end-to-end orchestrator due to its management capacity of all the different subsystems, the creation of end-to-end network services and the built-in communication framework with other datacenters. OSM also facilitates a plugin framework, allowing to use a variety of different software solutions as well as the inclusion of in-house, ready-to-use resource orchestration and VIMs. These plugins can be classified depending on which components of OSM they are integrated with, namely Resource Orchestrator (RO), User Interface (UI) and Service Orchestrator (SO). The RO is responsible for creating and placement of compute and networking resource, for interfacing with a SDN controller and it also allows the management of new VIMs and the resource coordination across multiple VIMs; the UI provides the user interface to the OSM orchestrator and the SO is responsible for end-to-end service orchestration.

On the other hand, the VNF architecture includes the VIM component, which controls the NFV infrastructure, i.e., the totality of hardware and software components that build the environment where VNFs are deployed. The telecommunications community has recognized the potential of OpenStack and it is well established as a viable platform for NFV [12]. In this sense, we selected OpenStack to provide the foundations for the NFV architecture, which offers standard APIs between the NFV elements, the infrastructure and the user interfaces. As mentioned above, a SDN solution has been run to support dynamic traffic flows operations; OpenFlow SDN solution running OpenDaylight Controller is deployed, thus its wide use in the telco ecosystem, allowing the creation of slices in the transport network to satisfy different levels of QoS, through the differentiation of different flows, and enforcing the adequate forwarding rules in the SDN switches.

The Fig. 3 describes the cloud orchestration architecture considered in this work, which allows to manage multiple datacenters. OSM is used to orchestrate VNFs across different VIMs (e.g., OpenStack and vendor's Cloud Execution Environment (CEE)), in this case the corresponding plugin, for the proprietary VIM CEE, had to be implemented. A customized OpenDaylight was also implemented to provide two different functionalities: flow control to enable intelligent networking managed by OSM and real-time monitoring of both infrastructure and network resources. Additionally, an UI plugin was deployed to present real-time information of network status (e.g., latency, jitter, throughput and others). The sections below will provide a detailed description of these implementations.

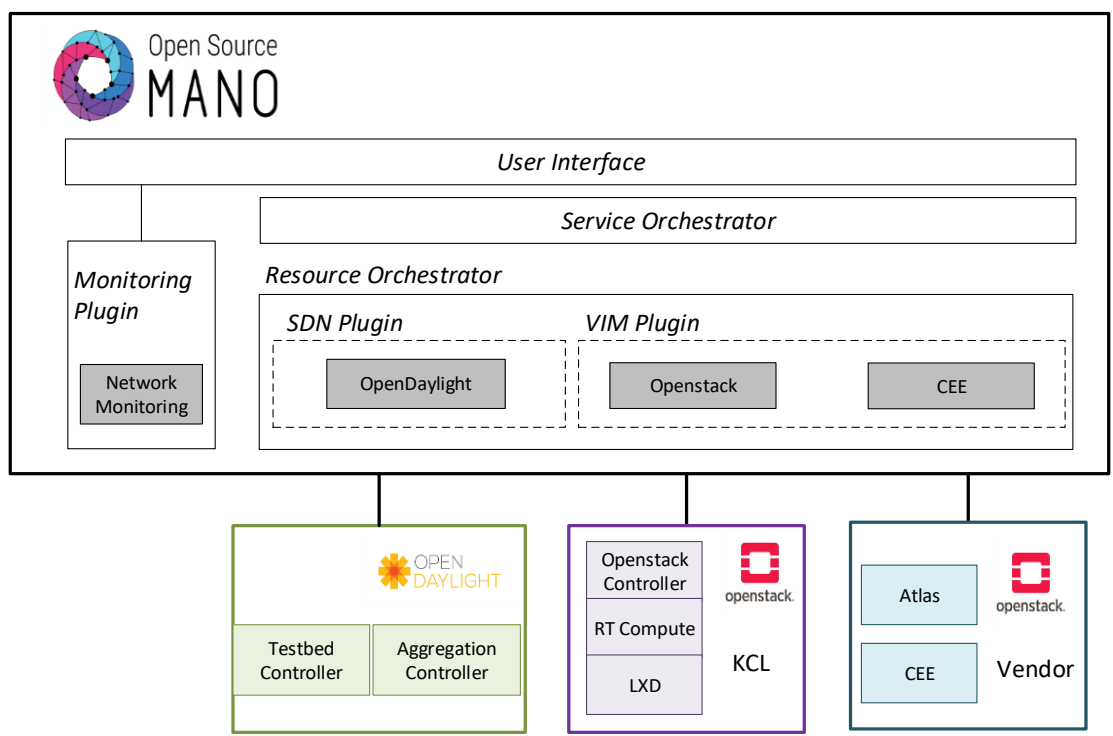

*Figure 3: Cloud Orchestration Architecture including OSM and VIM interfaces, and illustration of the developed plugins.*

### 3.1 Multi VIM Orchestration

In order to provide a scalable and flexible VIM orchestration, clouds from different vendors should be able to coexist within the same management system. In our network deployment, the systems integrated are a combination of OpenStack and vendor proprietary systems, such as the CEE, which is based on OpenStack but has added features to meet telecoms application requirements. Thus, to successfully integrate multiple clouds into a common orchestration platform, the necessary Application Programming Interfaces (APIs) to talk to the different VIMs needed to be implemented.

Building upon the plugin framework provided in OSM, multiple VIMs can be orchestrated and managed, as long as the appropriate VIM plugins are created and integrated within the RO. Currently, OSM support the following VIMs: OpenVIM, OpenStack, VMware vCloud Director, VMware Integrated OpenStack and Amazon Web Services.

For the KCL OpenStack, we use the built-in OSM interface mechanisms; however, for the vendor's system, we designed a new plugin that only provides access to a small subset of the APIs. Since any vendor may not wish to allow full access to their system's APIs when they participate in an orchestrated multi-vendor deployment, the OSM plugin framework, in this case, allows us to customize how the orchestrator accesses the CEE. In particular, access restrictions prevented direct connectivity between OSM and the CEE. This approach offers us the flexibility of manipulating multiple OpenStacks with a different level of access for each one and a different subset of VNFs and functions performed per VIM. It also shows how the API communication can be customized in OSM and how it is possible to integrate new APIs for interfacing with other platforms that are not currently supported.

The VIM plugin for the CEE was derived from the generic OSM-OpenStack plugin. Several modifications were made as the CEE authentication process differed from the standard OSM-OpenStack authentication process. To effectively authenticate the VIM plugin with the CEE, several components of the RO software had to be modified to properly pass unauthenticated requests to the CEE. To this end, to orchestrate seamlessly across all sub-systems, the corresponding APIs have been implemented, as well as the necessary integration components within OSM to successfully add the VIM.

The development of VIM specific plugins inherently leads to a compartmentalized orchestration framework. Compartmentalization provides administrators with the ability to specifically tailor functionality to VIMs (i.e., KCL-OpenStack and the CEE) as necessary. As discussed above, compartmentalization also reinforces security within the architecture as end users are only provided access to specific functions at the discretion of the VIM plugin developer and system administrator. Consequently, the overall architecture is inherently more scalable since additional functionality can be added to each VIM plugin as required.

To preserve the integrity of the larger system, both 4G and 5G systems required the necessary translations to be implemented and embedded into the VIM plugin. In doing so, the upper orchestration layer remained agnostic to the implementation details of the subsystem without any loss in the orchestration of non-heterogeneous sub-systems.

In terms of VNF's orchestration, OSM requires a communication interface with each VIM for the deployment of VNFs and optionally needs to deploy the configuration primitives in the VNFs for each network service via a direct connection to the management interface of the VNFs. Additionally, a set of management policies in the VIM ensure the desired fine-tuning compute resource allocation in cases where OSM does not offer the configuration options in its API. Each VIM plugin exposes northbound interfaces defining available network, storage and compute capability as well as per-tenant usage and billing information. The VIMs southbound interface collects and sorts this information into a Network Service Catalogue which details the available network services. Each service found in the catalogue is the composition of a set of information elements which define the lifetime, allocated CPU's and storage requirements for the instantiation process, namely VNF Descriptors.

## 3.2 Foundations for Smart Orchestration

One of the most important contributions in the management and orchestration community is the capability of real-time monitoring of both infrastructure and network resources. Monitoring enables the efficient use of resources, both in the physical infrastructure in charge of compute and storage of network services and the underlying transport network [13], which is often overprovisioned to avoid poor performances in the case of congestion.

Monitoring tools in OSM are at very early development stage, still experimental and only retrieve basic information from each VIM [14]. However, its modular nature allows to integrate custom mechanisms, as depicted in Fig. 4. The one developed, interfaces OSM with a customized OpenDaylight controller capable of measuring latency, jitter, packet loss, throughput and bandwidth across the network. This is achieved by using a Server capable of requesting and storing network data from many sources including OpenDaylight and OpenStack but any other data sources can be added easily with a simple REST API. It also allows applications other than OSM to use this data and build on top of it.

Extendibility and security are the key points of designing the monitoring framework as its not limited to OpenDaylight or OpenStack but any SDN controller and VIM can be interfaced. Every external application gets a token upon signup which has a certain expiry time and this must be passed in at every call for authentication. The Server can be seen as the Central Data Aggregation Point which runs many Microservices to collect network data from many places and store it in a timeseries fashion using InfluxDB, and allows other applications to use the data by providing both HTTP and HTTPS services and storing user details with MongoDB to facilitate the separation between management data from network data. Lastly, the Production Manager is responsible for monitoring the Microservices and restarting them upon crashing while keeping a log of all events, this is also integrated with Keymetrics UI which shows the physical state of the servers as well the state of Microservices. For monitoring the data in a graphical way, the Server is integrated with Grafana which is a tool for generating real-time graphs directly from InfluxDB.

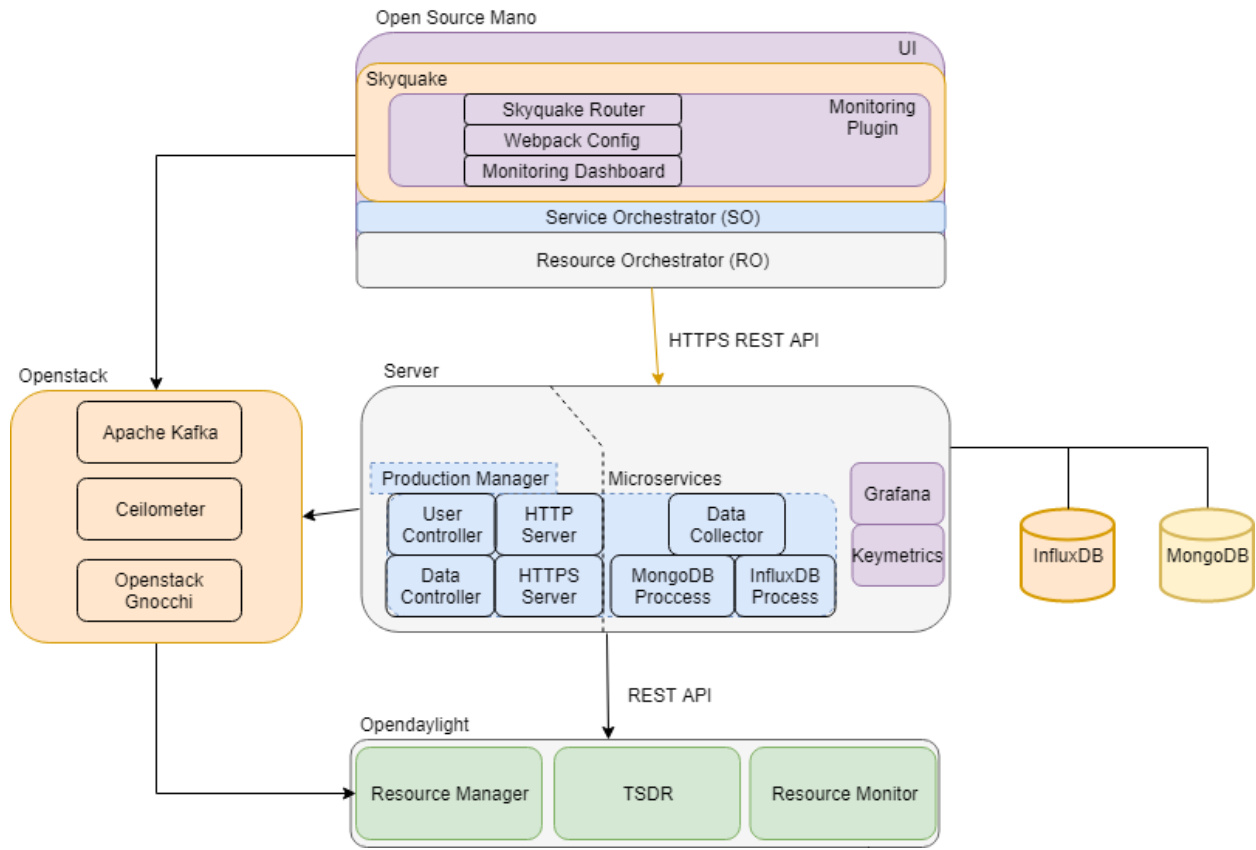

*Figure 4: Network Monitoring Plugin based on SDN Controller.*

The inclusion of such modules allows QoS policies such as prioritization and resource isolation through physical network slicing to be applied through the resource manager, which sits in the SDN controller. Development tools in OSM include mechanisms to expand the SDN integration and allow for automatic configuration of network forwarding rules and flows, according to the service requirements. In this way, the orchestration framework will ensure a correct coordination of policies and QoS enforcement rules, across the entire cloud infrastructure and physical network.

## 4. Cloud and IoT Deployment

To demonstrate the true potential of 5G, we have deployed in Central London a live 5G system, composed of 5G (pre-commercial) radio technologies, 4G and Wi-Fi capabilities. The 5G is based on two 3.5GHz Massive MIMO systems and a 28GHz mm-wave system. The radios are connected to a 5G basebands units, 5G networking equipment and 4G virtualized RAN (vRAN) and virtualized Core (vCore). Seven testbeds haven been interconnected, as Fig. 5 illustrates. The first is the outdoors 3.5GHz Massive MIMO and 28GHz mm-wave platform, on KCL's Strand campus rooftop. The second is indoors in the KCL 5G lab using a 4G eNB. The third uses 4G commercial picocells, using vendor 4G Dots. The fourth relies on KCL's software defined radio infrastructure, whereas the fifth uses Wi-Fi access points. The sixth is an outdoor test in the Borough of Westminster, and finally, the seventh is a live 5G scenario connecting KCL with the renowned London Guildhall. We aim to test each one independently against NGMN's 5G performance KPIs, and to test end-to-end communication capabilities across the different test sites by exploring new solutions for edge caching and real-time service orchestration based on network conditions, application requirements and user mobility.

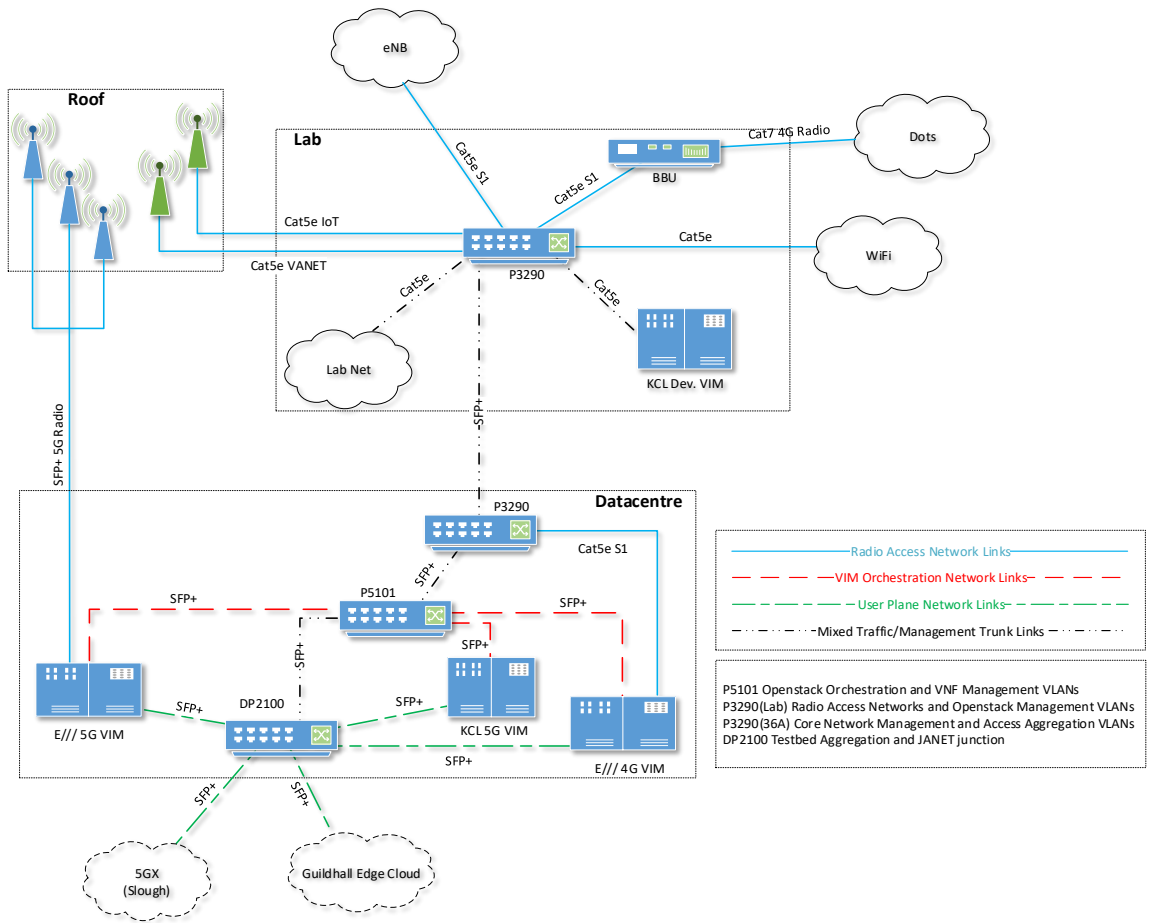

*Figure 5:* KCL's 5G cloud and integration system, composed of different air interface technologies (5G, 4G, Wi-Fi) and different virtualized RANs and Cores.

### 4.1 KCL's Cloud Deployment

To provide a more flexible and dynamic environment and to offer more options for experimentation, a set of features are desired in the final deployment: low latency applications support, pass through functionalities, hot-plug devices and NVF migration. In this section, we focus on the KCL-OpenStack (KCL 5G VIM, in Fig. 5) deployment and its characteristics. However, there are a total of six OpenStack systems in this 5G infrastructure: one hosting the KCL-OAI implementation and miscellaneous services (KCL 5G VIM), one VNF testing platform (KCL Dev. VIM), one 4G system (E/// 4G VIM), three for the core and the radio access (28 GHz and 3.5GHz) networks of the 5G system (E/// 5G VIM).

On one hand, the KCL 5G VIM deployment consists of eight compute nodes, out of which one is acting as the controller and seven are dedicated to compute and storage. The nodes are divided into two groups: container-based and virtual-machine hypervisors. Furthermore, one node from each group is running a low latency kernel for applications with execution time constraints such as 3GPP function splits. The nodes are also capable of Single-Root I/O Virtualization (SRIOV) which allows VNFs to pass through directly to a physical network interface to improve performance. The ability to migrate VNFs between hosts for edge-caching scenarios is guaranteed with a 10Gbps dedicated block storage network, where each compute host acts as a storage pool member.

The networking deployment exposes the VNFs to the physical infrastructure by means of 10Gbps links dedicated to the provider network and SRIOV. In this context, the OpenStack's component Neutron manages everything concerning to the networking VNF deployments inside OpenStack, while

OpenDaylight oversees switching and routing among the different testbeds outside OpenStack. Open vSwitch (OvS) is used on all nodes as the main bridge connecting VNFs to the physical networks when SRIOV is not used. For added performance, OvS with Data Plane Development Kit can also be utilized if desired for increased network performance and resource efficiency.

On the other hand, OpenDaylight is used as the provider network switch which manages a flexible switching and routing platform (based on DP2100 in Fig. 5), where all the VNFs are connected to the physical network. In the built-in integration of OSM and OpenDaylight, OSM takes full control of the SDN controller and installs static flows for the VNFs when they are instantiated. This scenario is known as SDN-assist and requires that the VNFs use SRIOV but becomes redundant in MAC-learning scenarios. To provide a more flexible and dynamic environment SDN-assist was rejected for this integration. OpenDaylight runs autonomously with L2-switch modules to perform MAC-learning function and forward packets. This gives us the ability to perform standard MAC-learning functions on the provider network and allows us to hot-plug devices as needed by the experimenters without having to redeploy Network Services with updated descriptors through OSM.

**4.2 5G-Enabled IoT Deployment**

One of our prime aims was to demonstrate the viability and usefulness of 5G technology in real-world applications. The main purpose is to demonstrate a fully immersive experience in the Guildhall spaces to promote the Cultural Hub of the City of London. The ultra-low latency technology yields a sensation of immediacy; and the wide bandwidth a sensation of immersion. For the specific purpose, KCL worked with the City of London, the Guildhall School of Music and Drama, and The Young People's Chorus of New York city to deliver an immersive cultural performance connecting performers across the ocean. Shown in Fig. 6 (left), we have thus installed the (pre-commercial) standardized 5G system onto the roof of KCL's Strand building, as well as an equivalent system in the Guildhall building in the City of London.

The 5G installation is underpinned by the following: first, we have network slicing enabled at both the virtualized core and transport network. Furthermore, edge-cloud computing for content and data aggregation at the edge is made possible through the ultra-low latency high-bandwidth system. This enables real-time big data analytics and real-time remote interactions.

Applications of the technology are manifold. It has attracted high interest in data collection for empowering city activities through advanced IoT technologies. For instance, it enables the analysis and display of geo-located financial data; or data analytics of real-time twitter data to be conducted.

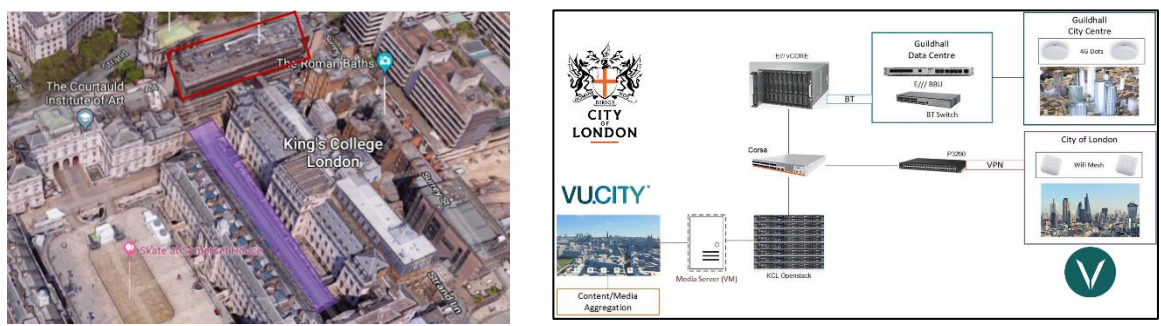

*Figure 6: Installation of 5G equipment onto the Strand roofs of KCL (left); and immersive 5G installation in London's iconic Guildhall (right).*

## 5. Conclusion and Future Outlook

The telecom community has been working on defining 5G to provide the latency, reliability and scalability needed for several critical services in the IoT space. The 5G implementation presented in this paper enables IoT deployments, based on a standardized architecture that allows the new real-time fully immersive applications (i.e., robotic tele-operation, force feedback haptic signal and virtual and augmented reality), where ultra-low latency is key and essential for vertical industries to provide these services within certain levels of QoS.

Moreover, we implemented an effective framework that provides management and orchestration mechanisms of multiple systems, being one of the open research challenges in IoT networks. In this context, virtualization and network slicing are a powerful enabler for 5G services, such as the Internet of Skills or Synchronized Reality [1,15]. Once fully operational, it will allow for a much more cost-efficient service provisioning and a substantial shift in the telco value chain. However, we observed with our designs and deployments that two transformations still need to happen. First, virtualization and slicing have to be enabled end-to-end beyond the telco networks; that is, the entire communication infrastructure globally must become software-enabled. Second, the orchestration and control frameworks need to handle the extremely large degree of freedom of the thus-created softwarized infrastructure; that is, artificial intelligence needs not only to advance significantly but also be trusted by the stakeholders deploying and using it.

**Acknowledgements**

The authors would like to thank the support of Ericsson which was one of the main vendors enabling the 5G deployments exposed in this paper. Furthermore, we are grateful for the financial support by DCMS UK Government for 5GUK, the EPSRC for Initiate, and the EC for 5GPPP Norma & VirtuWind.

**Bibliography**

[1] M. Dohler, T. Mahmoodi, M. A. Lema and M. Condoluci, "Future of mobile", 2017 European Conference on Networks and Communications (EuCNC), Oulu, 2017, pp. 1-5.

[2] B. Han, V. Gopalakrishnan, L. Ji and S. Lee, "Network function virtualization: Challenges and opportunities for innovations", in IEEE Communications Magazine, vol. 53, no. 2, pp. 90-97, Feb. 2015.

[3] R. Jain and S. Paul, "Network virtualization and software defined networking for cloud computing: a survey", in IEEE Communications Magazine, vol. 51, no. 11, pp. 24-31, November 2013.

[4] X. Sun and N. Ansari, "EdgeIoT: Mobile Edge Computing for the Internet of Things," in IEEE Communications Magazine, vol. 54, no. 12, pp. 22-29, December 2016.

[5] 3GPP TS36.802, "Evolved Universal Terrestrial Radio Access (E-UTRA); NB-IOT; Technical Report for BS and UE radio transmission and reception", Release 13, July 2016.

[6] G. Mountaser, M. L. Rosas, T. Mahmoodi and M. Dohler, "On the Feasibility of MAC and PHY Split in Cloud RAN", 2017 IEEE Wireless Communications and Networking Conference (WCNC), San Francisco, CA, 2017, pp. 1-6.

[7] 3GPP TR38.801, "Study on new radio access technology: Radio access architecture and interfaces", Release 14, March 2017.

[8] 3GPP TR23.501, "System Architecture for 5G", Release 15, December 2017.

[9] A. C. Baktir, A. Ozgovde and C. Ersoy, "How Can Edge Computing Benefit From Software-Defined Networking: A Survey, Use Cases, and Future Directions," in IEEE Communications Surveys & Tutorials, vol. 19, no. 4, pp. 2359-2391, Fourthquarter 2017.

[10] Broadband Forum, "TR-348 Hybrid Access Broadband Network Architecture", July 2016 [Online] Available at: https://www.broadband-forum.org/technical/download/TR-348.pdf

[11] 3GPP TR 23.793 "Study on Access Traffic Steering, Switch and Splitting support in the 5G system architecture", February 2018.

[12] OpenStack Foundation Report, "Accelerating NFV Delivery with OpenStack", White Paper, 2016. [Online] Available at: https://www.openstack.org/telecoms-and-nfv/

[13] Q. Chen, F. R. Yu, T. Huang, R. Xie, J. Liu and Y. Liu, "An Integrated Framework for Software Defined Networking, Caching, and Computing", in IEEE Network, vol. 31, no. 3, pp. 46-55, May/June 2017.

[14] ETSI OSM Community, "OSM Release Three – A Technical Overview", White Paper, October 2017. [Online] Available at: https://osm.etsi.org/images/OSM-Whitepaper-TechContent-ReleaseTHREE-FINAL.pdf

[15] Ericsson – King's College London innovation output, "Reshaping our world with 5G research", [Online] Available at: https://www.ericsson.com/en/networked-society/innovation/reliable-communications/kings-college

**Biographies**

**Idelkys Quintana-Ramirez** (idelkys.quitana@kcl.ac.uk)**:** received the BEng in Telecommunications and Electronics Engineering and the MSc in Telematics from Central University of Las Villas (Cuba) in 2004 and 2010, respectively. She is currently a PhD candidate at the Communications Technologies Group of the University of Zaragoza. Her research interests include network optimization and Quality of Experience in Multimedia Services (video streaming, P2P-TV applications and other real-time services), cloud orchestration of network services and SDN.

**Anthony Tsiopoulos** (anthony.tsiopoulos@kcl.ac.uk): received his MSc from the University of Sussex in Scientific Computation and Mathematics (2014) and an MSc in Computing and Security from King's College London (KCL) (2015) where he then continued his research in Cloud Technology, Telecommunications, Virtualization and Software Implementation in the KCL 5G Telecommunications Laboratory. His research interests include Security and Privacy, Distributed Networks and Software Implementation of Network Programming and Virtualization.

**Maria A Lema** (maria.lema_rosas@kcl.ac.uk): received her PhD degrees from the Universitat Politecnica de Catalunya in 2015. She is currently the lead researcher of the Ericsson-Sponsored 5G Tactile Internet Project with the Centre for Telecommunications Research, King's College London. She has been involved in the definition of use cases for 5G. Most of her research in this area has been focused on designing end to end ultra-reliable low latency 5G networks in the context of the Internet of Skills.

**Fragkiskos Sardis** (fragkiskos.sardis@kcl.ac.uk)**:** received his MSc in Computer Networks in 2010 and his PhD in Mobile Edge Computing in 2015. As a research associate at KCL, he leads the architecture design of the 5G testbed in collaboration with Ericsson in 2017 and he is currently the head of the 5G infrastructure at KCL. His interests include cloud orchestration, NFV, SDN, cloud robotics and vehicular networks.

**Luis Sequeira** ([luis.sequeira@kcl.ac.uk](mailto:luis.sequeira@kcl.ac.uk)): received his MSc in Digital Communications from University of Costa Rica in 2008, and M.Sc. in Mobile Networks from University of Zaragoza in 2012. He finished his PhD (Cum Laude) in Information Technologies at University of Zaragoza in 2015. His research interests focus on QoS in multimedia systems, analysis and modeling of access networks and multimedia services. Currently, he is a Research Associate at King's College London working on SDN and orchestration of network services.

**James Arias:** received an MSc from McMaster University (Canada) in Computational Science and Engineering (2015) with a research focus on High Performance Computing and Applied and Computational Mathematics. James began working with the 5G Telecomunnications Laboratory at Kings College London in 2017 as a software developer where he primarily focussed on the development of a user space implementation of Multipath-TCP.

**Aravindh Raman** (aravindh.raman@kcl.ac.uk): is a PhD candidate in the Department of Informatics at King's College London, prior to which he was a researcher at Indian Institute of Technology Delhi. His research centers on contributing to a robust future Internet infrastructure through network measurements and building next generation content delivery architectures. Aravindh holds a Bachelor's degree in Information Technology from Anna University, Chennai.

**Ali Azam** ([ali.azam@kcl.ac.uk](mailto:ali.azam@kcl.ac.uk)): is a 5G researcher/developer working on many projects at Kings College London including Virtuwind by European Commission and 5GUK project by UK government. He specialises in Software Defined Networks, Network Slicing, Virtual Infrastructure Management and 5G Orchestration. Ali has worked with many leading firms including Ericsson, NEC, Siemens, Intel and many more.

**Mischa Dohler** ([mischa.dohler@kcl.ac.uk](mailto:mischa.dohler@kcl.ac.uk)): is currently a Full Professor of Wireless Communications with the King's College London and a Co-Founder of the pioneering smart city company Worldsensing. He is a fellow of the Royal Society of Arts and a Distinguished Member of Harvard Square Leaders Excellence. He is a frequent keynote, panel, and tutorial speaker. He has pioneered several research fields, contributed to numerous wireless broadband, IoT/M2M, and cybersecurity standards, holds a dozen patents, organized, and chaired numerous conferences, has over 200 publications, and authored several books. He acts as a Policy, Technology, and Entrepreneurship Adviser. He is also an Entrepreneur, a Composer, and a Pianist with four albums on iTunes, and fluent in six languages. He has talked at TEDx. He had coverage by national and international TV & radio, and his contributions have featured on the BBC and the Wall Street Journal.